# Text Mining-Based Patent Analysis for Automated Rule Checking in AEC

Zhe Zheng, Bo-Rui Kang, Qi-Tian Yuan, Yu-Cheng Zhou, Xin-Zheng Lu, and Jia-Rui Lin*

Department of Civil Engineering, Tsinghua University, Beijing, China, 100084.
`lin611@tsinghua.edu.cn`

**Abstract.** Automated rule checking (ARC), which is expected to promote the efficiency of the compliance checking process in the architecture, engineering, and construction (AEC) industry, is gaining increasing attention. Throwing light on the ARC application hotspots and forecasting its trends are useful to the related research and drive innovations. Therefore, this study takes the patents from the database of the Derwent Innovations Index database (DII) and China national knowledge infrastructure (CNKI) as data sources and then carried out a three-step analysis including (1) quantitative characteristics (i.e., annual distribution analysis) of patents, (2) identification of ARC topics using a latent Dirichlet allocation (LDA) and, (3) SNA-based co-occurrence analysis of ARC topics. The results show that the research hotspots and trends of Chinese and English patents are different. The contributions of this study have three aspects: (1) an approach to a comprehensive analysis of patents by integrating multiple text mining methods (i.e., SNA and LDA) is introduced ; (2) the application hotspots and development trends of ARC are reviewed based on patent analysis; and (3) a signpost for technological development and innovation of ARC is provided.

## 1    Introduction

Authorities' rules such as design guidelines, codes, standards, and international, national, or local authorities' laws stipulate the safety, sustainability, and comfort of the entire lifecycle of a built environment [1]. For a long time, the regulatory compliance of a building design was checked manually by domain experts. However, this process is not only immensely time-consuming but is also challenging for project participants [2] due to the complexity and ambiguity of rules and regulations [3]. Therefore, automated rule checking (ARC, also known as ACC) has been extensively studied by many researchers for nearly 70 years [4], and several rule-based ARC systems have been established [5]. For example, CORENET (short for COnstruction and Real Estate NETwork), which is the first large effort of ARC, was initiated in 1995 in Singapore [6]. A universal circulation network (UCN) is developed as a plug-in based on the Solibri model checker (SMC) to check occupant circulation rules of the US court design guide [7]. Other rule-based ARC systems include BCAider [8], Jotne EDModelChecker [9], and so on. However, the above rule-based ARC systems are costly to maintain and inflexible to modify and are often referred to as black-box approaches [10]. Hence, to make the process more flexible and transparent, attempts have been made to (1) automated interpretation of the rules based on natural language processing (NLP) [11, 12, 13], (2) automated alignment of BIMs and computer-interpretable regulations [14], and so on.

To promote future development in the field, many efforts have been devoted to the analysis of the development trends. For example, Eastman et al. [5] examined and surveyed five rule checking systems that assess building designs according to various criteria in detail. Ismail et al. [15] reviewed the previous studies, which successfully employed the appropriate techniques in interpreting the rules for checking purposes. Fuchs [16] assessed the state-of-the-art of NLP for building code interpretation by analyzing 42 research articles published since 2000. These works have pointed out the key research areas and analyzed the current and ongoing research and development (R&D) directions. However, the existing studies mainly focused on the existing ARC systems or research articles, while seldom considering the patent analysis. Patent analysis is valuable to throw light on the innovation and development of a technological field over the course of time [17, 18].

Therefore, this work aims to explore the application hotspots and trends of compliance checking by analyzing the related patents based on text mining methods. The patents from the database of the Derwent Innovations Index database (DII) and China national knowledge infrastructure (CNKI) are taken as data sources. Then the analysis is comprised of three steps. First, the quantitative characteristics and annual distribution analysis of patents are carried out. Second, several topics of the ARC patents are identified using a latent Dirichlet allocation (LDA). Third, to reveal the relation between different topics, the identified ARC topics co-occurrence analysis is carried out based on Social Network Analysis (SNA). Finally, based on the analysis results, the application hotspots and trends are revealed. The findings are expected to be valuable for the related stakeholders including researchers, ARC system developers, and companies. The introduced approaches to the analysis of ARC patents are also useful for other scenarios.

## 2 Methodology

This study utilizes patents as data resources and integrates several text mining technologies including the LDA topic identification technology, and the SNA topic correlation analysis to identify the application hotspots of ARC and forecast its trends. Fig. 1 shows the outline and workflow of this research. Meanwhile, the toolkits used to implement the workflow are also displayed at the bottom of Fig. 1. The methods deployed are detailly illustrated from Section 2.1 to Section 2.3.

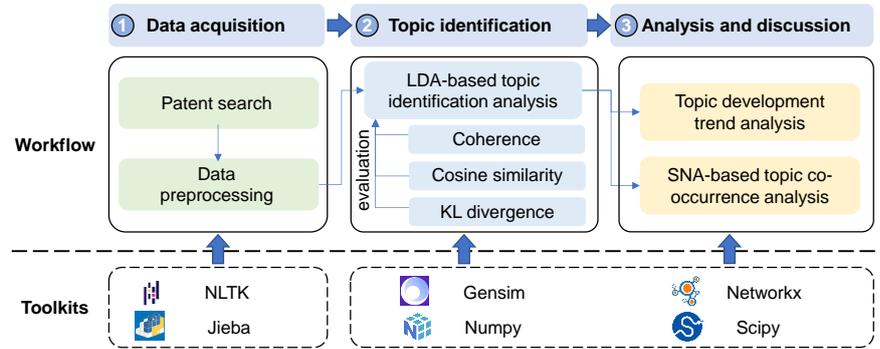

**Fig.1.** Workflow of this research

## 2.1    Data acquisition

In this work, the patent retrieval process consists of (1) screening by keywords and, (2) screening manually by domain experts to filter the patents that are not relevant to the rule checking. The Chinese and English patents are obtained from the CNKI and DII database, respectively, and the retrieval time is from 2010 to 2021. For the Chinese patents, the keyword is "design review", and after the manual screening, total 113 Chinese patents were collected for analysis. For the patents in English, the keywords include "rule checking", "compliance checking", "design review", "automated compliance checking". After the manual screening, finally, total 121 patents in English were collected for analysis.

The abstracts of the collected patents are used for further analysis. After the patent retrieval process, the data preprocessing consists of tokenization, special characters (e.g., !%$#& *?,/.;"\) removal, and stop words removal. For the Chinese patents, the original texts are tokenized and removed stop words using the Python lib named Jieba, while the English patents are processed using the Python lib named NLTK (Natural Language Toolkit).

## 2.2    Topic identification

After the acquisition and pre-processing of the patent data, the topic identification is performed to mine the latent information of patents. Topic identification methods (i.e., topic modeling methods) are typical text mining methods. The Latent Dirichlet Allocation (LDA) model proposed by Blei et al. [19] is one of the most widely-used topic modeling methods [18, 20, 21, 22]. LDA model is a generative probabilistic model of corpus where each document is represented as random mix-tures over latent topics, and each topic is characterized by a distribution over words [19]. The basic idea of the LDA model is shown in Fig. 2. The hyperparameter $\alpha$ is the Dirichlet prior on the document-topic distributions and the hyperparameter $\beta$ is the Dirichlet prior on the topic-word distributions. LDA model assumes the following steps to generate a document [18, 19].

1. For each document $\boldsymbol{d}$, sample a topic proportion $\theta_d$ from the Dirichlet distribution $Dir(\alpha)$.

2. For each word $w_{d,n}$ to be generated in the document $\boldsymbol{d}$:
   a. Choose a topic $z_{d,n} \sim Multinomial(\theta_d)$
   b. Choose a word $w_{d,n} \sim Multinomial(\varphi_k \mid k = z_{d,n})$, where the word proportion per topic $\varphi_k$ is sampled from the Dirichlet distribution $Dir(\beta)$.

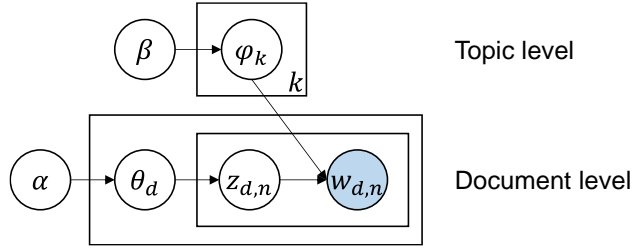

**Fig. 2.** Latent Dirichlet Allocation (LDA) model

To determine the optimal number of topics generated by the LDA model, the coherence scores of the topics are used [24]. The coherence metric uses statistics and probabilities extracted from the reference corpus to evaluate how well a topic is. The larger the coherence scores are, the better the effect of LDA topic identification. The coherence measure method consists of four main steps: (1) segmentation, (2) probability calculation, (3) confirmation measure, and (4) aggregation [24]. The coherence model in the genism toolkit is used to implement the evaluation process.

Besides, to evaluate the performance of the topic generation results, the topic distances among the topics are measured via Kullback–Leibler (KL) divergence and cosine similarity analysis. According to the LDA model, each topic is represented as a set of words with proportional. So, based on KL divergence and cosine similarity, the distance matrix should be constructed, where the element in the $i$ row and the $j$ column represents the specific distance value between the number $i$ topic and the number $j$ topic. For the KL divergence analysis, the distance matrix is an asymmetric matrix. The larger the value of the non-diagonal elements of the distance matrix, the greater the differentiation of words between two topics and the better performance of topics generated by the LDA model. For the cosine similarity analysis, the distance matrix is a symmetric matrix. The smaller the value of the non-diagonal elements of the distance matrix, the greater the differentiation of words between two topics and the better performance of topics generated by the LDA model.

### 2.3 Topic co-occurrence analysis

After the LDA analysis, the topic of each patent is recognized. To further explore the topic relevance between the identified topics, the SNA-based topic relevance analysis is performed. SNA is often used to explore the publication, citation, and cocitation networks, collaboration structures, and other forms of social interaction networks [25].

In this work, the nodes in the topic relevance graph are the identified topics. Meanwhile, the link between two nodes means there is at least one article that contains the two topics at the same time. According to the LDA model, each article consists of a bag of topics, where the larger the percentage of topics, the more important the topic is. Therefore, when a topic accounts for more than 10 %, we consider this topic to be the article's main topic. If an article has multiple main topics at the same time, these topics are called co-occurrence topics. Then a link will be generated for these two topics. The attribute of each link means the number of articles that contains the two topics. Therefore, the topic relevance graph is a typical undirected symmetric network. After the construction of the topic relevance graph, Gephi [26] is then used to perform the visualization of network analysis.

## 3 Results

### 3.1 Quantitative characteristics of ARC patents

The annual publication numbers of patents can reflect the development trend and research interests. Therefore, after the patent data acquirement and preprocessing, the trend analysis is conducted, as shown in Fig. 3. The first Chinese patent on rule checking appeared in 2011, while the first English patent appeared in 2013. The trend of the Chinese and English patents is similar. The number of patents grows slowly until 2015. From 2016 to 2021, the growth rate of the related patents has significantly accelerated, which indicates that the rule checking methods and systems are gaining more and more interest in recent years.

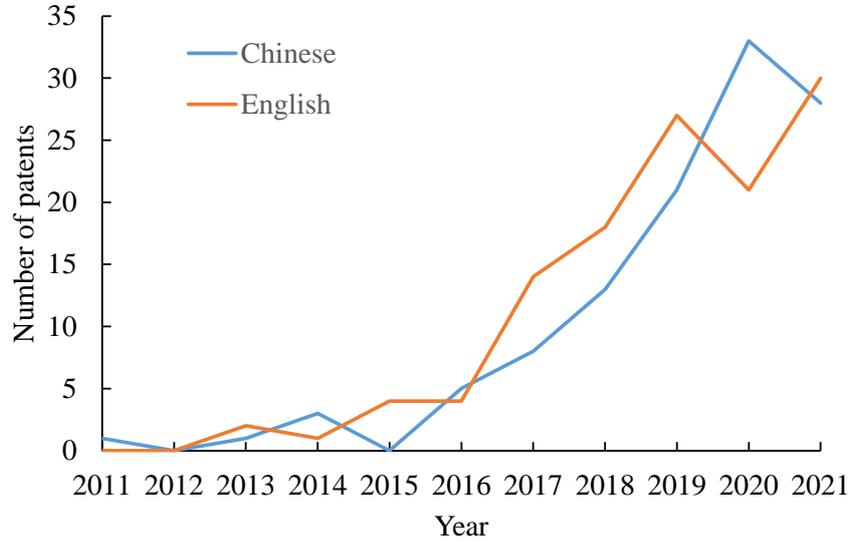

**Fig. 3.** The number of patents on rule checking

### 3.2 LDA-based topic clustering of ARC patents

To further analyze the application hotpots and development trends of the ARC patents, the LDA model is utilized to explore the main topics and their corresponding keywords from the collected ARC-related patents. First, the relatively proper topic numbers are searched, as shown in Fig. 4. The coherence values are used to measure the rationality of the topics. The best topic number is 5 and 7 for the Chinese patents and English patents, respectively. The coherence value of the Chinese topics is higher than that of the English topics, indicating that the quality of Chinese topics is higher than that of English topics.

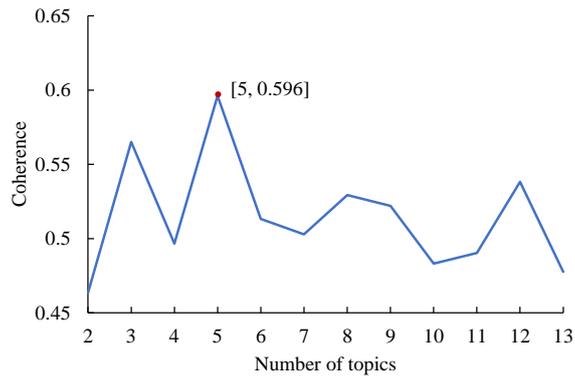

(a) Chinese patents

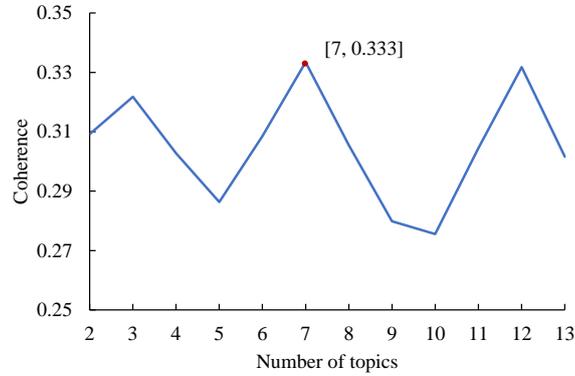

(b)English patents

**Fig. 4.** Coherence values of the topics constructed by LDA

To further evaluate the performance of the topic generation results, the topic distances among the topics are calculated via Kullback–Leibler (KL) divergence and cosine similarity analysis. The distance matrixes of the topics are shown in Fig. 5. For the Chinese patents, as shown in Fig. 5a, $A_{13}$ ($A_{ij}$ means the element at the number $i$ row and the number $j$ column) and $A_{25}$ have the largest cosine similarity. $A_{13}$ and $A_{25}$ have the smallest KL divergence, as shown in Fig. 5b. The results of the two matrixes are consistent, indicating that topic 2 and topic 5 have the highest similarity, and topic 3 and topic 2 have the second-highest similarity. For the English patents, as shown in Fig. 5c, $A_{25}$, $A_{27}$, $A_{56}$, and $A_{67}$ have the largest cosine similarity. $A_{52}$ and $A_{76}$ have the smallest KL divergence, as shown in Fig. 5d. The results of the two matrixes are consistent, indicating that the similarity between topics 2 and 5 and the similarity between topics 6 and 7 are relatively high. In general, the similarity between Chinese topics is lower than that between English topics, indicating that the quality of Chinese topics is higher than that of English topics. This is consistent with the result of coherence analysis, as shown in Fig. 4.

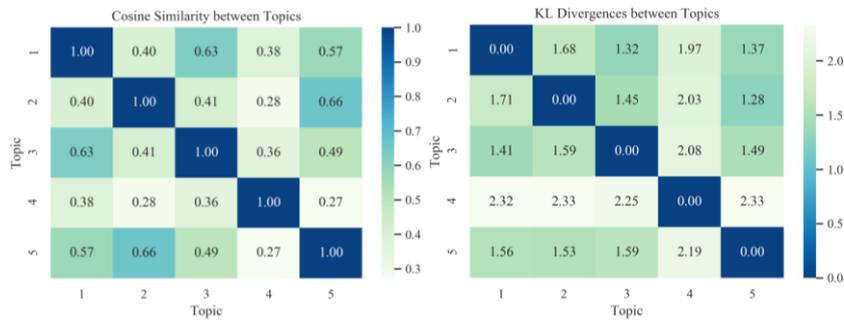

(a) Cosine similarity for Chinese patents    (b) KL divergence for Chinese patents

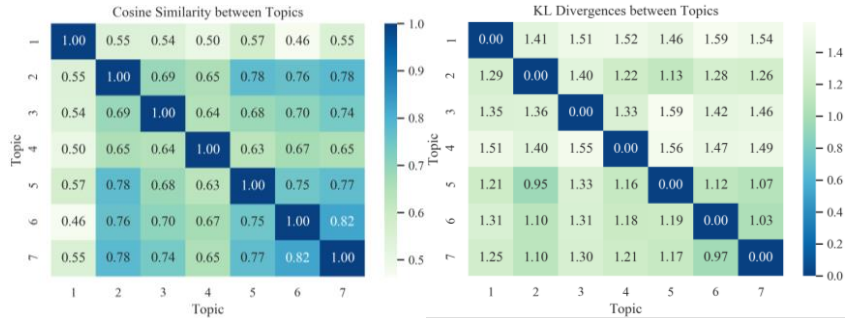

(c) Cosine similarity for English patents    (d) KL divergence for English patents

**Fig. 5.** Distance matrixes of the topics

Then all the patents are analyzed, and the identified topics according to the word distributions are listed in Tables 1 and 2. The top 10 words and their corresponding weight values (percentage) are also listed in Tables 1 and 2. For the Chinese patents, the topics are topic 1 "codes database and auxiliary software", topic 2 "drawings and models information extraction", topic 3 "CAD element identification modules", topic 4 "User interaction", and topic 5 "Auxiliary devices and systems". Topics 2 and 5 may share some keywords on drawings or model information; therefore, the similarity is relatively high.

For the English patents, the topics are topic 1 "digital system and terminal", topic 2 " BIM applications in design and construction", topic 3 " data integration and management", topic 4 "computer-executable rule and knowledge base", topic 5 "BIM-based management platform", topic 6 "Auxiliary tools for rule checking", and topic 7 "Rule Checking systems". Topic 2 and topic 5 are about BIM-related applications; therefore, the similarity between the two topics is relatively high. And topics 6 and 7 are about auxiliary tools or systems for rule checking; consequently, the similarity between the two topics is also relatively high, which is consistent with the results of Fig. 5c and 5d.

**Table 1.** Distributions of the topic of the Chinese ARC-related patents

| Topics | Keywords and probabilities | | | | | | | | | |
|---|---|---|---|---|---|---|---|---|---|---|
| Codes data-base and auxiliary software | Design review | Rule check | module | codes | target | system | data | database | method | region |
| | 2.7 | 2.7 | 2.3 | 1.5 | 1.4 | 1.2 | 1.1 | 1.1 | 1.0 | 1.0 |
| Drawings and models' information extraction | draw-ings | check | component | Design review | building | method | infor-mation | applica-tion | target | acquire-ment |
| | 7.7 | 2.8 | 2.8 | 2.6 | 1.8 | 1.5 | 1.3 | 1.3 | 1.2 | 1.1 |
| CAD ele-ment identi-fication modules | data | Design review | infor-mation | module | Construc-tion draw-ings | model | Rule check | terminal | method | CAD |
| | 2.7 | 2.3 | 1.8 | 1.6 | 1.4 | 1.3 | 1.3 | 1.3 | 1.2 | 1.2 |
| User interac-tion | Design review | inter-face | display | design | change | Phase diagram | method | model | files of drawings | click |
| | 3.6 | 2.9 | 2.6 | 2.5 | 1.7 | 1.7 | 1.4 | 1.4 | 1.4 | 1.2 |
| Auxiliary devices and systems | draw-ings | target | Design review | infor-mation | module | build-ing | project | compo-nent | applica-tion | detection |
| | 3.1 | 2.6 | 2.3 | 1.7 | 1.6 | 1.5 | 1.3 | 1.2 | 1.1 | 1.1 |

Note: The numbers in the table represent the percentage of words in the topic.

**Table 2.** Distributions of the topic of the English ARC-related patents

| Topics | Keywords and probabilities | | | | | | | | | |
|---|---|---|---|---|---|---|---|---|---|---|
| Digital | design | digital | infor- | drawing | building | review | delivery | terminal | modeling | description |

| | | | | | | | | | |
|---|---|---|---|---|---|---|---|---|---|
| system and terminal | | mation | | | | | | | |
| 3.3 | 2.9 | 2.9 | 2.8 | 1.8 | 1.5 | 1.3 | 1.3 | 1.2 | 0.9 |
| BIM applications in design and construction | model | information | BIM | drawing | software | building | construction | design | rtm | 3D |
| 3.7 | 2.5 | 2.0 | 1.8 | 1.3 | 1.3 | 1.3 | 1.2 | 1.2 | 1.2 |
| Data integration and management | BIM | information | model | drawing | building | system | based | object | data | unit |
| 2.6 | 2.4 | 1.9 | 1.8 | 1.7 | 1.5 | 1.2 | 1.2 | 1.1 | 0.9 |
| Computer-executable rule and knowledge base | drawing | BIM | model | module | substation | system | building | review | information | arch |
| 2.1 | 1.9 | 1.9 | 1.8 | 1.4 | 1.4 | 1.2 | 1.2 | 1.1 | 1.1 |
| BIM-based management platform | construction | drawing | BIM | model | design | information | data | project | according | management |
| 3.3 | 2.8 | 2.6 | 2.5 | 2.0 | 1.7 | 1.4 | 1.3 | 1.2 | 1.2 |
| Auxiliary tools for rule checking | model | BIM | rule | drawing | component | design | construction | description | building | set |
| 4.0 | 3.8 | 2.4 | 2.3 | 1.6 | 1.3 | 1.2 | 1.0 | 0.9 | 0.9 |
| Rule Checking systems | model | BIM | component | information | data | building | drawing | design | construction | check |
| 3.5 | 2.7 | 2.2 | 2.1 | 2.1 | 2.0 | 1.8 | 1.6 | 1.4 | 1.0 |

Note: The numbers in the table represent the percentage of words in the topic.

### 3.3　Analysis and discussion

After the topic construction via the LDA models, the topic-based quantification analysis is conducted to explore the trends, and the SNA-based topic co-occurrence analysis is performed to provide more insight into the combination and development of the technologies.

As shown in Fig.6a, for the Chinese patents, with the development of information technology (e.g., natural language processing, deep learning), the patents related to ARC have shown an increasing trend since 2016, and have grown substantially since 2018. Topic 2 "Drawings and models' information extraction" has the highest growth rate and quickly became the most published topic. The result indicates that researchers are more concerned about how to extract and integrate the information from drawings and models. Topic 1 "Codes database and auxiliary software" has received sustained attention in recent years. Topic 4 "User interaction" and topic 5 "Auxiliary devices and systems" have also grown rapidly since 2018. The result shows that good user interaction methods and more convenient auxiliary tools are important to ARC systems. Topic 4 and topic 5 started late and developed rapidly, which shows that making the ARC system more convenient may be the future research direction.

As shown in Fig.6b, for the English patents, the patents related to ARC have shown an increasing trend since 2016. Topic 5 "BIM-based management platform", topic 6 "Auxiliary tools for rule checking", and topic 7 "Rule Checking systems" grew faster and were published the most. This shows that many researchers pay attention to the development of platforms and systems. Among them, topic 5 contains the most patents because it's about the whole life cycle of BIM-based management where automated rule checking may be used in different steps. Topic 4 "Computer-executable rule and knowledge base" currently receives relatively little attention. However, Ismail pointed out that rule interpretation is the most vital and complex stage in the rule checking process [15]. To achieve a fully automated rule checking system, more attention may be paid to the topic 4 related research. Topic 2 "BIM applications in design and construction" has a downward trend, which may be a result of many studies not only focusing on the design stage and the construction stage but also focusing on the whole life cycle of the buildings.

It can be concluded from Fig.6a & 6b that the content of Chinese patents and English patents are different. Chinese patents pay more attention to the various components of the ARC system. In contrast, the English patents focus more on integrating the ARC system in various BIM-based applications.

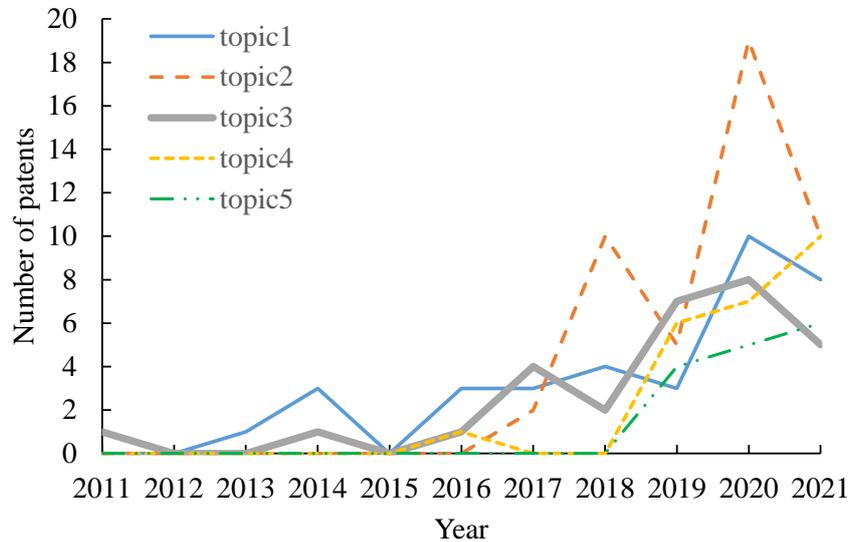

(a) Chinese patents

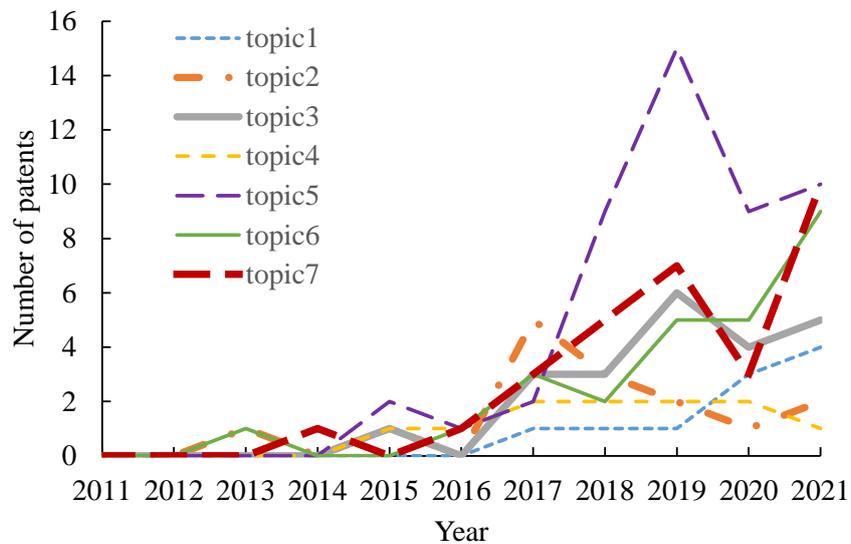

(b)English patents

**Fig. 6.** The number of patents on different topics constructed by LDA

For the co-occurrence analysis, the results are shown in Fig. 7. For the Chinese patents, topic 1 "Codes database and auxiliary software" and topic 2 "Drawings and models information extraction" have the most co-occurrences. The results show that when the patent involves drawing and model information extraction, it will also include other auxiliary tools. For the English patents, topic 5 "BIM-based management platform" co-occurs with topic 3 "Data integration and management", topic 6 "Auxiliary tools for rule checking", and topic 7 "Rule Checking systems", respectively. Because topic 5 mainly includes a full life cycle building management platform, which involves data integration and other auxiliary tools.

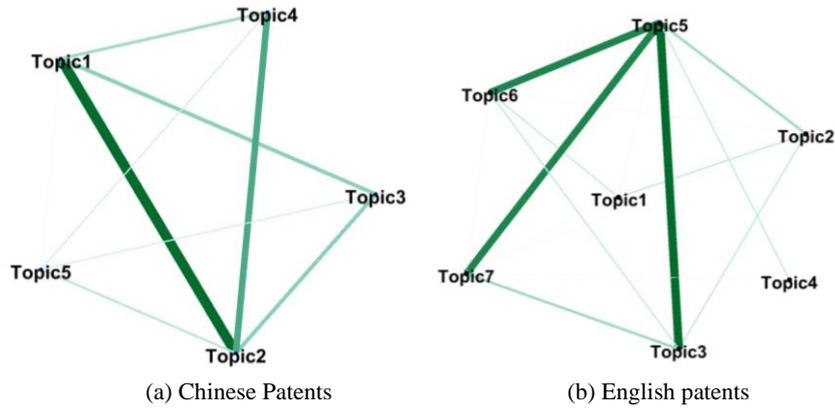

(a) Chinese Patents                    (b) English patents

**Fig. 7.** Co-occurrence of the topics

## 4    Conclusions

ARC is expected to promote the efficiency of the compliance checking process in the AEC industry. To analyze the trend development and hotpots of the ARC, this paper first extracted the patents' texts from the database of Derwent Innovations Index database (DII) and China national knowledge infrastructure (CNKI). Subsequently, a three-step analysis approach is carried out, which includes (1) quantitative characteristics (i.e., annual distribution analysis) of patents, (2) identification of ARC topics using a latent Dirichlet allocation (LDA) and, (3) SNA-based co-occurrence analysis of ARC topics. The conclusions are as follows:

(1) The trend of the Chinese and English patents is similar. The number of patents grows slowly until 2015. From 2016 to 2021, the growth rate of the related patents has significantly accelerated.

(2) The topic models are identified based on the LDA model, and the quality of the constructed topics has been measured via three methods, including coherence value, cosine similarity, and Kullback–Leibler (KL) divergence. The results of the three methods are consistent.

(3) For the Chinese patents, the patents related to "drawings and models information extraction" are gaining more attention. Besides, the results show that the number of patents that focus on improving the user experience of the ARC system has grown rapidly. Therefore, making the ARC system more convenient may be the future research direction. For the English patents, the patents related to " BIM-based management platform" are gaining more attention. While the patents related to "Computer-executable rule and knowledge base " received relatively little attention currently. The rule interpretation is one of the most vital stages in the rule checking process. Future research can pay more attention to the rule interpretation to achieve a fully automated rule checking system.

The limitations of this study are listed as follows:

(1) The performance of LDA analysis can be further improved for better topic detection. Besides, more methods should be utilized to validate the results of LDA.

(2) Only the abstracts of the patents are used to perform the analysis, more structured and unstructured data from patents is expected to use in the future.

### Acknowledgment

The authors are grateful for the financial support received from the National Natural Science Foundation of China (No. 51908323, No. 72091512), the National Key R&D Program (No. 2019YFE0112800), and the Tencent Foundation through the XPLORER PRIZE.